\documentclass[amsmath,amsfonts,amssymb,footnotes]{article}
\usepackage{ulem}
\usepackage{slashed}
\usepackage{color}

\begin{document}

\title{Quantum Degenerate Systems}
\author{Fiorenza de Micheli$^{\dag 1,2}$ and Jorge Zanelli$^{\ddag 1,3}$\\
$^1$  {\small \textit{Centro de Estudios Cient\'{\i}ficos, Arturo Prat 514, Valdivia, Chile.}} \\
$^2$ {\small \textit{Instituto de F\'{\i}sica, Pontificia Universidad Cat\'olica de Valpara\'{\i}so,  Casilla 4059, Valpara\'{\i}so, Chile.}} \\ 
$^3$  {\small \textit{Universidad Andr\'es Bello,  Av. Rep\'ublica 440, Santiago, Chile.}}\\
 \texttt{$^{\dag}${\small zuccavuota-at-gmail.com}, $^{\ddag}${\small z-at-cecs.cl}}}
\maketitle

\begin{abstract}
Degenerate dynamical systems are characterized by symplectic structures whose rank is not constant throughout phase space. Their phase spaces are divided into causally disconnected, nonoverlapping regions such that there are no classical orbits connecting two different regions. Here the question of whether this classical disconnectedness survives quantization is addressed. Our conclusion is that in \textit{irreducible} degenerate systems --in which the degeneracy cannot be eliminated by redefining variables in the action--, the disconnectedness is maintained in the quantum theory: there is no quantum tunnelling across degeneracy surfaces. This shows that the degeneracy surfaces are boundaries separating distinct physical systems, not only classically, but in the quantum realm as well. The relevance of this feature for gravitation and Chern-Simons theories in higher dimensions cannot be overstated. 
\end{abstract}


\maketitle            
\section{Introduction} 
Classical degenerate systems are characterized by an evolution which is globally determined by the equations of motion, except on a certain set $\Sigma$ of measure zero in phase space $\Gamma$. On this set the evolution is indeterminate because the matrix that multiplies the highest derivatives in the evolution equations --the Hessian matrix in the Lagrangian formalism or the symplectic form in the Hamiltonian description--, degenerates: its determinant vanishes there \cite{STrZ}.

Many physically relevant systems including gravitation and Chern-Simons theories in dimensions greater than four \cite{TZ}, vortex interactions in fluids \cite{Aref,Choodnovsky}, and piecewise smooth systems in electronics, mechanics or engineering (the so-called Filippov systems), exhibit this feature \cite{Filippov}.

Generically, the degeneracy sets are surfaces of codimension one in phase space, and higher codimension at their intersections. These surfaces split the phase space into nonoverlapping regions, causally disconnected from each other, each describing a nondegenerate system. A classical degenerate system would lose degrees of freedom irreversibly if its orbits reach a degeneracy surface; the Liouville current is not conserved at the degeneracy surfaces where there is a net ingoing or outgoing flux. The sign of the flux distinguishes between sources and endpoints of orbits. In the latter case, once the system reaches the domain wall, it generically acquires a new gauge invariance and one degree of freedom is frozen, while the remaining degrees of freedom evolve regularly thereafter \cite{STrZ}.

Degeneracies of this kind present challenges that require extending the standard treatments. For instance, in Dirac's time-honored approach for constrained Hamiltonian systems \cite{Dirac}, the rank of the symplectic form is less than the phase space dimension but constant, while in degenerate systems the rank is not constant throughout phase space. Some degenerate systems like, Chern-Simons theories in dimensions higher than three, may also have a Dirac matrix of second class constraints whose rank in not constant throughout phase space (irregularity), which makes some second class constraints become first class on some submanifolds of phase space \cite{MZ}. This problem makes the formulation of a quantum problem much more complex and will not be discussed here.

Although degenerate systems could be viewed as extensions of constrained systems, degeneracy is explicitly excluded from the hypotheses of the standard Dirac approach, and this introduces conceptual difficulties that must be addressed \cite{Gitman-Tyutin}. Our aim is to clarify to what extent the difficulties in degenerate systems are insurmountable obstacles for their quantization, or whether they can be circumvented, reducing the problem to one already known and solved. Our conclusion is that degenerate systems can be quantized following the standard postulates of quantum mechanics, although they exhibit a number of peculiar and unexpected features.

\section{Classical degenerate systems} 
In this section we analyze the classical dynamics of degenerate systems, building on the ideas developed in~\cite{STrZ}. In order to fix ideas, let us consider a dynamical system described by the first order action
\begin{equation}
I[z;1,2]=\int_{t_1}^{t_2} [A_i(z) \dot{z}^i +A_0(z)]dt, \quad \mbox{with     }  i =1,2,\cdots 2n,
\label{azione}
\end{equation}
This can be viewed as an action in Hamiltonian form, where $z^i$ are noncanonical coordinates in phase space $\Gamma$. Alternatively, $L=A_i \dot{z}^i+A_0$ describes the Lagrangian for a system that has $2n$ primary (second class) constraints $\phi_i=p_i-A_i(z)\approx 0$. In spite of its simplicity, this system captures the problem and, at the same time, describes
any  Lagrangian system (degenerate or not) of a finite number of degrees of freedom \cite{BsAs}. 

\subsection{Review of degenerate systems} 
In this subsection, we review the results reported in \cite{STrZ}. The equations of motion read
\begin{equation}
F_{ij}\dot{z}^j+E_i=0,  \label{eq_motion}
\end{equation}
where\footnote{Here we are interested in autonomous systems with time-independent $A_i$, but this could be easily generalized to include the time-dependent case, where $E_{i} = \partial_{i} A_{0} - \partial_{0} A_{i}$. }
\begin{equation} 
F_{ij} \equiv\partial_i A_j - \partial_j A_i, \quad \mbox{and} \quad E_i \equiv \partial_i A_0. \label{eq_motion_appendix}
\end{equation}
This defines a Hamiltonian system where the pre-symplectic two-form $F=dA$ is defined by the skew-symmetric $2n\times2n$ matrix $F_{ij}$. From (\ref{eq_motion}) it follows that $\dot{z}^iE_i\equiv 0$, and therefore the orbits are contained in the surfaces $A_0=$const. Let us summarize the basic facts about degenerate systems.  

$\bullet$ {\bf Degeneracy surfaces}. Solving (\ref{eq_motion}) for $z^{i}(t)$ requires inverting $F_{ij}(z)$, which is not defined on the set 
\begin{equation}
\Sigma=\{ z \in \Gamma \, | \, \Delta=0 \}\, ,
\end{equation}
where $\Delta(z)=\det [F_{ij}(z)]$ vanishes. The degeneracy set $\Sigma \subset \Gamma$ is defined by one relation among the coordinates $z^i$, and it therefore generically corresponds to a collection of codimension one surfaces, which divide the phase space into disconnected regions.\\

$\bullet$ {\bf Causally disconnected dynamics}. From the equations of motion (\ref{eq_motion}), it is clear that if $F_{ij}$ has simple zeros, the sign of $\dot{z}$ changes at $\Sigma$. This means that the classical evolution cannot take the system across the degeneracy surfaces: there is no causal connection between states separated by a degeneracy surface. The question that naturally arises is whether this condition continues to hold if quantum mechanics is included. Can there be tunnelling across $\Sigma$? What happens to a wave packet prepared on one side that corresponds to a classical trajectory that approaches
$\Sigma$? This will be discussed in Section 4.\\

$\bullet$ {\bf Robustness of degeneracies}. As every skew-symmetric matrix, $F_{i,j}$ can be block-diagonalized by a local orthogonal transformation, 
\begin{equation}
F_{ij}= \left[ \begin{array}{ccc}
  \begin{array}{cc}
  0 & f_1 \\\
-f_1 & 0 \end{array} &    &     \\
     &  \begin{array}{cc}
  0 & f_2 \\\
-f_2 & 0 \end{array} &   \\
    &     & \ddots
\end{array}
\right] . \label{block_diag}
\end{equation}
In an open, simply connected set where $\Delta(z)=[f_1(z)f_2(z)\cdots f_n(z)]^2>0$, a coordinate redefinition would allow to put this pre-symplectic form into the canonical symplectic form, in which $f_1(z)=f_2(z)=\cdots=f_n(z)=1$ (Darboux theorem). Clearly, this is not possible for degenerate systems in  an open set containing points of degeneracy, $z\in \Sigma$, where at
least one of the $f_r$ vanishes.

The degenerate character of a dynamical system is a feature that cannot be eliminated by an appropriate change of coordinates. To see this, it suffices to observe that the determinant of the pre-symplectic form, $\Delta=$det$(F_{ij})$, transforms as a pseudoscalar under coordinate changes
\begin{equation}
\Delta\rightarrow \Delta'=J^2 \Delta\, ,
\end{equation}
where $J$ is the Jacobian of the transformation $z\rightarrow z'$. Hence, the zeros of $\det F$ cannot be removed by general coordinate transformations in phase space, unless one is willing to accept singular coordinate transformations. Darboux's theorem, however, continues to hold within every nondegenerate domain, where local coordinates can be chosen such that $f_i(z)=1\, \forall i$ (see Appendix). \\

$\bullet$ {\bf Intrinsic two-dimensionality of degeneracies}. Since $F$ is exact it is necessarily closed, $dF=0$, which implies that each $f_r(z)$ is a function of a unique pair of coordinates,
\begin{equation}
F=\sum_{r=1}^n f_r(z^{2r-1},z^{2r}) \, dz^{2r-1} \wedge dz^{2r} \, ,
\end{equation}
Hence, the zero's of $\Delta(z)$ describe analytically the degeneracy surfaces in the two-dimensional surfaces spanned by the coordinate lines $z^{2r-1}\,,\,z^{2r}$. This means that in order to analyze the dynamical properties of degenerates systems, it is sufficient to focus on two-dimensional surfaces embedded in phase space. In particular the equations of motion (\ref{eq_motion}) appear as a system of $n$ decoupled equations of two variables $(z^1,z^2)$, which depend parametrically on the remaining coordinates $z^a$,
\begin{equation}
\left\{\begin{array}{cc}
 f(z^1,z^2)\,\dot{z}^1 &= - \partial_2 A_0 (z^1,z^2; z^a)  \\
-f(z^1,z^2) \,\dot{z}^2 &= -\partial_1 A_0 (z^1,z^2; z^a) \; .
\end{array}
 \right.\label{genericsystem}
\end{equation}
\\
$\bullet$ {\bf Degenerate dynamic flow}. In \cite{STrZ} $f$ was assumed to be a Morse function, so the zeros can be generically assumed to be simple, and the velocity must change sign at the degenerate surfaces. This means that the orbits either start, end, or run tangent to the set $\Sigma$. Moreover, if $f(z^1,z^2)$ is a Morse function, $\Sigma$ can be generically either an infinitely long line or a closed curve in the $(z^1-\,z^2)$-plane, forming a collection of Jordan curves. Exceptionally, $\Sigma$ may have a finite number of self intersections and isolated points or cusps which can be removed by a continuous deformation of $A_0$.

Moreover, the velocity field $\dot{z}^i$ has a nonvanishing divergence,
\begin{equation}
\partial_i \dot{z}^i =f^{-2} \epsilon^{ij}\partial_i f \partial_j A_0 \neq 0 \, .
\end{equation}
Consequently, the time evolution of a degenerate system does not preserve the volume in phase space: the volume $v$ of a small region in phase space evolves as
\begin{equation}
\mbox{div} \, v (z) = f^{-2} \, \Big\{f , H \Big\}\, ,
\end{equation}
which blows up as the orbit approaches a degeneracy point: $v \rightarrow \pm \infty$, depending on the sign of the gradient of $f$ along the orbit.

\subsection{Dynamical role of the degeneracy}    

The system (\ref{genericsystem}) represents a $2$-dimensional vector field, not necessarily smooth, but mildly singular due to the unbounded velocity $\dot{z}\rightarrow \infty$ at $f(z^1,z^2)= 0$. This system represents a \textit{continuous directional field}\footnote{A direction field is called continous if $tg(\alpha)$ depends continously on the points $({z}^{1},{z}^{2})$.} given by 
\begin{equation}
{\dot{z}^{2}}/{\dot{z}^{1}} := \tan\alpha(z^1,z^2,z^a)=-\partial_1A_0(z)/\partial_2 A_0(z), \nonumber
\end{equation}
whose integral curves are completely determined by $A_0(z)$. This expression is insensitive to the change $t\rightarrow -t$, so it carries no information about the reversal of orientation that takes place at the points where the orbits intersect the degenerate surfaces $\Sigma$. More importantly, this expression is also independent of $f$, and therefore, there are infinitely many dynamical systems analogous to (\ref{genericsystem}), whose orbits have the same shape but with different dynamics and different degenerate surfaces (in particular, $f(z^1,z^2)=1$ gives the simplest nondegenerate analogous system, corresponding
to a standard mechanical system, with Hamiltonian $H=-A_0$, and $z^1=p$ and $z^2=q$).

The level curves $A_0 (z^1,z^2;z^a) = constant$ implicitly define the shape of the orbits, while $f=\partial_1 A_2(z^1,z^2) - \partial_2 A_1(z^1,z^2)$ determines the dynamics, i.e. the pace at which the orbits are traced. In other words, the Hamiltonian draws the orbits and the pre-symplectic form determines the time evolution.

\section{Reducible and irreducible degenerate actions}   

We have seen that coordinates can be found such that the equations of motion take a canonical form everywhere within a nondegenerate domain, and don't seem to have any problem; but, can those equations be obtained from an action principle?  Can the dynamical system within a nondegenerate domain be described by a regular, \textit{nondegenerate action principle}?  Can the action of a degenerate system like (\ref{azione}) be replaced by a nondegenerate action that reproduces the same evolution within a region that does not contain degeneracies?

The question is whether any of the infinitely many equivalent nondegenerate descriptions can be obtained from an action principle of the type (\ref{azione}). As we will see next, the answer is negative, as stated in the following

\textbf{Lemma:} Given a \textit{generic} degenerate system obtained from the action principle $I[z]$ as in (\ref{azione}), none of its nondegenerate analogues can be obtained from a local action principle $\tilde{I}[z]$. The only exceptional (non-generic) case in which an action principle exists for both, degenerate and non-degenerate systems, occurs if the degeneracy function $f(z)$ is a constant of motion, or equivalently, if the orbits do not intersect the surface $f(z)=0$.

\textbf{Proof}: Suppose there exists an action $\tilde{I}[z]$, for which the equations of motion are nondegenerate,
\begin{equation}
\dot{z}^i  = \epsilon^{ij}\,\frac{\partial \tilde{A}_0(z)}{\partial z^j}\, .
\end{equation}
Since these equations describe the same degenerate orbits as described, for instance, in (\ref{degsystem}) of the appendix, then
\begin{equation}
\partial_i \tilde{A}_0 (z) = f^{-1} \partial_i  A_0 (z)\, .
\end{equation}
A fast check of the mixed second partial derivatives shows that
\begin{equation}
\vec{\nabla} f   = \varphi  \vec{\nabla}{A_{0}} \label{grad_f}\, .
\end{equation}
where $\varphi$ is any scalar function, or equivalently,
\begin{equation}
\epsilon^{ij} \partial_i f \partial_j A_0=\big\{ f, A_0\big\} = 0\, ,
\label{f_constant}
\end{equation}
where $\{ \cdots , \cdots \}$ is the Poisson bracket. Eq. (\ref{grad_f}) means that the level curves of $f$  and $A_0$ must concide, and (\ref{f_constant}) implies that $f(z)$ is a constant of motion.  In other words, only the action for degenerate systems whose orbits run tangent to the degenerate surfaces can be replaced by an action describing a nondegenerate system; this type of degeneracy is a \textbf{reducible} one. \textbf{Irreducible} degenerate systems instead are those that cannot be described by an equivalent nondegenerate action principle, whose classical orbits intersect the degenerate surfaces.

For example, a degenerate system 
\begin{eqnarray}
f(z)\epsilon_{ab} \dot{z}^b = E_a(z)
\end{eqnarray}
is reducible iff $\partial_a f = \varphi(z) E_a(z)$; otherwise, it is irreducible. This conclusion is relevant for the study of quantum degenerate systems. The point is that, in order to discuss the quantum mechanics of a particular system, it is not sufficient to have its dynamical equations, it is necessary to know the action principle that defines it \cite{Hojman-Shepley}. As is well known,
systems without an action principle --like a damped harmonic oscillator-- do not have a well defined quantum mechanical description.\\

\section{The quantum problem}  

As we have seen, irreducible degenerate system cannot be obtained from a non-degenerate action principle. This means that the quantization of irreducible degenerate systems is a problem that cannot be addressed following the standard procedures of quantum mechanics as it applies to nondegenerate systems. The peculiar feature is that the Dirac bracket not only depends on the coordinates, but moreover, it diverges on the degenerate surface.  When the symplectic structure degenerates and is no longer invertible, what is the correct approach to define the quantum theory?

There are two standard constructions of a quantum theory starting from a classical one: the path integral and the canonical (Schr\"odinger) approach. Here, we analyze the simplest irreducible degenerate system following the second one. 

Let us consider a generic 2-dimensional first order Lagrangian of the form
\begin{equation}
L(x,y)=A_{x} \dot{x} + A_{y} \dot{y} + A_{0} \, .
\end{equation}
\noindent
The Dirac bracket is given by the inverse of symplectic form,
\begin{equation}
\{x,y\}^{*}=\frac{1}{f(x,y)} \,  ,
\label{deg_diracbracket}
\end{equation}
where $f=\partial_{x}A_{y}-\partial_{y}A_{x}$.

The phase-space coordinates have a noncanonical symplectic structure, and there is no metric and no preferred coordinate system in the problem. However, since for irreducible degenerate systems, $f$ has been assumed to be a smooth Morse function, whose level curves do not coincide with the level curves of $A_0$, a natural option would be to take the value of $f$ as a coordinate, which may be called $x$. The level curves of Morse functions are either closed or infinitely extended, and the coordinate lines for $x$ can be identified with the gradient of $f$ in a local patch.

Quantization still requires finding an adequate prescription of operators such that the Dirac bracket (\ref{deg_diracbracket}) becomes the commutator at the quantum level,
\begin{equation}
\left[ \hat{x},\hat{y}\right] =i \hbar \frac{1}{x}\, . 
\label{commutator}
\end{equation}
Then $x$ and $y$ can be promoted to operators satisfying this commutation relation, 
\begin{eqnarray}
\hat{x} &:&=x  \label{deg_prescription} \\
\hat{y} &:&=-i \hbar \frac{1}{x}\partial _{x}\, .
\label{operator_prescriptions}
\end{eqnarray}
The quantum operator $\hat{H}=\hat{H}(\hat{x},\hat{y})$ that replaces the classical Hamiltonian $H_{c}=-A_{0}(x,y)$, is a singular differential operator with a leading coefficient $1/x$. Still, given the fact that classically the energy is conserved, $\dot{H}_{c}=0$, we expect the quantum Hamiltonian to have observable eigenvalues, and therefore $\hat{H}$ must be self-adjoint, which eventually depend on the correct choice of boundary conditions that define the Hilbert space. \\

\subsection{Example: The simplest first order Lagrangian} 
 
We illustrate the procedure by analyzing the simplest Lagrangian discussed in~\cite{STrZ}: $A_x=xy$, $A_y=0$, $A_0=-\nu y$, $F_{ij}(x) =x\epsilon_{ij}$, so that
\begin{eqnarray}
L &=&xy\dot{x}-\nu y\, , \\   
H &=&\nu y=-A_{0} \, , \label{simplehamiltonian} \
\end{eqnarray}
whose degenerate surface at $x=0$ can be thought of as an approximation near the degenerate surface of a generic system. In spite of its simplicity --and possibly unrealistic nature--, this problem has some interesting features that help understand more general cases.
 
The classical solution is given by \cite{STrZ}
\begin{equation}
x^2(t) = 2 \nu t + x_{0} \, .
\end{equation} 
For $\nu<0$, the system presents an attractive surface of degeneracy at $x=0$ (repulsive for $\nu>0$). Note that the orbits flow towards this surface, reaching the degeneracy with infinite velocity in a finite time. Conversely, for $\nu <0$, the orbits emerge from this surface with infinite velocity and go to $\pm \infty$ with a velocity that approaches zero at infinity.

Using the above prescription (\ref{operator_prescriptions}) the Hamiltonian operator in this case is
\begin{equation}
\hat{H} = \nu \hat{y} =-i \hbar \nu \, \frac{1}{x} \partial _{x}\,  .
\label{hamiltonian}
\end{equation}
The domain of this singular differential operator must be chosen so that  the corresponding Hilbert space will be equipped with a well-defined weighted scalar product. In general the weighted Hilbert space $L^{2}(\Omega \subset \mathbf{R}, w(x)\,\mathrm{d} x) $\footnote{Also denoted as $L^{2}(\Omega \subset \mathbf{R},w(x))$ or $L^{2}_{w}(\Omega \subset \mathbf{R})$.} consists of (all equivalence classes of) complex-valued functions, defined on a subset $\Omega$ of $\mathbf{R}$, that are square-integrable with a weight $w(x)$,
\begin{equation}
 \Vert \varphi(x) \Vert = \Big{(} \int    \vert \varphi(x) \vert^{2}  w(x)\,\mathrm{d} x  \Big{)}^{\frac{1}{2}}  .  
\end{equation}

The weight $w(x)$ is chosen in such a way that the Hamiltonian (\ref{hamiltonian}) is symmetric,
\begin{equation}
\int \varphi_1^*(x) \left[\hat{H} \varphi_2(x)\right] w(x) dx = \int \left[\hat{H} \varphi_1(x)\right]^*\varphi_2(x)  w(x) dx\, ,
\end{equation}
up to boundary terms.  The symmetry condition plus the positivity of the scalar product require $w(x) = \vert F_{ij}\vert = \vert x \vert$. This is the measure implied by the noncanonical Dirac bracket (\ref{deg_diracbracket}), and is consistent with the presence of the degenerate surface at $x=0$. Hence, the domain where $\hat{H}$ defines a proper scalar product is
 \begin{equation}
\mathcal{D}_{o}(\hat{H}) = \{\psi \in L^{2}(\mathbf{R},|x| dx) : \hat{H}(\psi(x)) \in L^{2}(\mathbf{R}, |x| \, dx)\} .  \label{DoH} 
\end{equation}
The corresponding scalar product and norm in the Hilbert space are
\begin{equation}
<\varphi_1,  \varphi_2> = \int \varphi_1^{*} |x| \varphi_2 \,\mathrm{d} x,  \qquad  \mbox{and} \qquad
||\varphi || = \Big{(} \int  |\varphi |^{2}  |x| \, dx \Big{)}^{\frac{1}{2}}\, , 
\label{scalar_product}  
\end{equation}
respectively.

In this case, the Schr\"odinger equation reads
\begin{equation}
- i \hbar \nu \frac{1}{x} \frac{\partial}{\partial x} \Psi(x,t) =  i \hbar \frac{\partial}{\partial t} \Psi(x,t) , \label{eq_sch}
\end{equation}
which is a singular differential equation with indefinite weight $x$ that can change sign and vanish. \footnote{As an example, also Sturm-Liouville equations, $ -\frac{d}{dx} \big[ p(x) \frac{d y}{dx}\big] + q(x)\cdot y = \lambda w(x) \cdot y $, are second order singular equation with a weight $w(x)$. They are usually considered as a self-adjoint operators in the Hilbert space $ L^{2}(\Omega \subset \mathbf{R},w(x)\,\mathrm{d} x) $, but usually considering only intervals for which $w(x)$ has constant sign and $w(x) \neq 0$.}

The general solution of equation (\ref{eq_sch}) is $\Psi=\varphi(x^2-2\nu t)$, where $\varphi$ is any differentiable function. Since the classical system is conservative, the quantum states $\psi$ can be spanned in a basis of eigenstates of the Hamiltonian (\ref{hamiltonian}).  Hence, a stationary solution $\psi(x,t)$ is also an eigenstate of $\hat{H}$ of the form
\begin{equation}
\Psi_E(x,t)=\psi_{E}(x) \alpha_{E}(t)
\end{equation}
\noindent
with
\begin{equation}
\psi_E(x)= \psi_0 \,  e^{\frac{i}{\hbar } \frac{E}{2\nu} x ^{2} } \quad \mbox{and} \quad \alpha_E (t) = \alpha_0 \, e^{-\frac{i}{\hbar }E t} , \label{immediatesolution}
\end{equation}
where $E$= constant is an eigenvalue of $\hat{H}$.

The crucial point now is the choice of the domain where the Hamiltonian operator is self-adjoint. Let's stress that for unbounded (linear) operator, as $\hat{H}$, self-adjointness and symmetry may not coincide depending on the domain. In practice, the process to establish the self-adjointness requires the symmetry condition (see Appendix).  In our case, $\hat{H} = -i \hbar \nu \, x^{-1} \partial _{x}, $ is self-adjoint provided the functions in the Hilbert space $\mathcal{D}_{o}(\hat{H})$ satisfy appropriate boundary conditions, depending on whether the domain includes the degeneracy or not.

\subsection{Dealing with the degeneracy}    

In the presence of the divergence at $x=0$, the Schr\"{o}dinger equation (\ref{eq_sch}) can be solved by either restricting the domain so as to exclude the origin, or by imposing some additional boundary conditions involving the values of $\psi$ at $x=0^{\pm}$.  The latter option is a subtle issue in view of the first order nature of equation (\ref{eq_sch}).

\subsubsection{Excluding the degeneracy: $x \in (0,a)$}   

A simple possibility is to consider the domain $(0,a)$, in which case, the normalized stationary states are
\begin{equation}
\Psi_{\mbox{\tiny E}}(x,t)= \frac{\sqrt{2}}{a} \exp \left[ \frac{iE}{2\hbar \nu}  (x^2 -2\nu t) \right] . \label{wave(0,a)}
\end{equation}

This solution is even in $x$ and never vanishes in the range, although its domain of definition does not include $x \leq 0$. The equation is separable and therefore, the solution can be factorized as $\Psi(x,t)=\psi(x) \alpha(t)$. The symmetry condition
\begin{equation}
 <\hat{H} \psi, \phi> = <\psi,\hat{H} \phi > \quad \forall \psi, \phi \in \mathcal{D}(\hat{H}), 
\end{equation}
\noindent
reduces to
\begin{equation}
\psi^{*}(a)\phi(a) - \psi^{*}(0^{+})\phi (0^{+})=0.
\label{BC(0,a)}
\end{equation} 
It then follows that the operator $\hat{H}$ is self-adjoint in the space of functions which differ by an arbitrary but fixed phase $\theta$ at the end points of the domain, $\psi(a)=e^{i \theta}\psi(0^{+}) \neq 0$,  
\begin{eqnarray}
<\psi,\hat{H} \phi > = <\hat{H}^{\dag} \psi, \phi>,  \nonumber \\ 
\forall \psi,  \phi \in  \mathcal{D}_{(0,a)}(\hat{H}) \equiv \mathcal{D}_{(0,a)}(\hat{H^{\dag}}) \, ,
 \end{eqnarray}
where
\begin{equation}
\mathcal{D}_{(0,a)}(\hat{H}) = \{ L^{2}((0,a),\vert x \vert d x) :  \psi(a)=e^{i \theta}\psi(0^{+}) \neq 0 \} . \label{domain(0,a)}
\end{equation}
Hence, the eigenfunctions
\begin{equation}
\psi_{\mbox{\tiny E}}(x)= \frac{\sqrt{2}}{a} \exp \left[ \frac{iE}{2\hbar \nu} x ^{2} \right] \label{eigenfunction(0,a)}
\end{equation}
\noindent
form a complete orthonormal set spanning the space $\mathcal{D}_{(0,a)}(\hat{H})$.

The boundary condition in (\ref{domain(0,a)}) implies a discretization of the energy spectrum,
\begin{equation}
E_{n} := \frac{2 \nu \hbar}{a^2} (2 n \pi + \theta) = \frac{4 \pi \nu \hbar n}{a^2} + \frac{2 \nu \hbar}{a^2} \theta, \quad n \in \mathbf{Z}  \quad \mbox{and} \quad \theta \in [0, 2\pi]
\label{discretizationWn}
\end{equation}
\begin{equation}
E_{n} - E_{m}:= \Delta E = \frac{4 \pi \nu \hbar}{a^2} (n-m)
\end{equation}
 
 Thus, the energy eigenstates are described by the wave functions
\begin{equation}
\Psi_{n, \theta}(x,t)= \frac{\sqrt{2}}{a} \exp \left[ i \frac{2 n \pi +\theta}{a^2} ( x^2 - 2\nu t) \right]  ,
\end{equation}
and the general solution in the interval $(0,a)$ is $\Psi (x,t)=\sum c_{n} \Psi_{n, \theta}(x,t)$, with the coefficients $c_{n}$ given by
\begin{equation}
c_{n} = < \Psi_{n, \theta}(x,t) \vert \Psi (x,t)> = \int_{0}^{a}  \Psi_{n, \theta}^{*}(x,t) \Psi (x,t)  x d x
\end{equation}
which are determined by the initial condition $\Psi(x,0)=\psi_{0}(x)$.

The probability of finding the state in a configuration around $x$ is given by 
\begin{equation}
P(x<x'<x+d x,t)= |\Psi(x,t) |^2 |x| dx, \label{def_probability}
\end{equation}
which is the same for all $n$ and any value of $\theta$,
\begin{equation}
P_{n,\theta}(x<x'<x+dx,t) = |\Psi_{n,\theta}(x,t) |^2 |x| dx = \frac{2}{a^2} |x|  dx\, . \nonumber
\end{equation}

The parameter $\theta$ must be the same for all the functions in the Hilbert space. Different choices of $\theta$ give rise to different Hilbert spaces which, however, describe equivalent physical systems.\footnote{The parameter $\theta$ produces a shift of energy levels by the constant $\Delta E=2 \nu \hbar\theta/a^2$, which can be seen as the effect of putting the system in an
environment at a constant potential \cite{Nobili}.} Hence, changing the value of $\theta$ has no effect on the energy differences between states, or on the matrix elements $<\Psi_1(x,t) \mathbf{M} \Psi_2(x,t)>$ for any operator $\mathbf{M}$ (e.g., in the probability amplitude). Without loss of generality, $\theta$ can be set to zero, which implies the added symmetry $\Psi_n^{*} =
\Psi_{-n}$ among the energy eigenstates.

The energy spectrum $E_n$ is unbounded below because the are no restrictions on the values of $n \in \mathbf{Z}$ as seen in (\ref{discretizationWn}).  This is due to the first-order character of the Schr\"odinger operator in this case. Thus, the spectrum of this Schr\"odinger equation is analogous to the Dirac case, where the negative energy states are interpreted as anti-particles states going backwards in time. Here, the energy eigenvalues $E_{n}$ remain unchanged under simultaneous reversal of $n$ and $\nu$,
\begin{equation}
 E_{n,\nu} = -E_{-n,\nu} = -E_{n,-\nu}= E_{-n,-\nu}\, .
\end{equation}
Hence, the negative values of $E$ correspond to the states of a system where $\nu$ has the opposite sign, passing from a system of attractive character to a repulsive one or vice-versa, which are precisely the time-reversal of each other.

\subsubsection{Including the degeneracy: $x\in (a^-,a^+)$, with $a^-<0<a^+$}   

Let us now solve the Schr\"odinger equation (\ref{eq_sch}) in an interval that spans across the surface of degeneracy. The idea is to describe a situation in which both the initial and the final states can be on either side of $x=0$, in order to explore the possibility of tunnelling across the degeneracy surface. The difficulty is not so much to find the space of solutions in the interval
$(a^-,a^+)$, with $a^-<0<a^+$, ---in other words, a Hilbert space $L^{2}((a^-,a^+), |x| dx)$--- but to make sure that $\hat{H}$ is self-adjoint in that space of solutions. In fact, the solution analogous to the previous case, normalized in this domain with measure $|x|dx$, is
\begin{equation}
\psi_E(x)= \sqrt{\frac{2}{(a^-)^2+(a^+)^2}} \, \exp \left[i\frac{E}{2\hbar \nu} x^2 \right]\, ,
\label{ansatz9(1a,b)}
\end{equation}
which reduces to (\ref{wave(0,a)}) for $a^-=0$, $a^+=a$. The condition (\ref{BC(0,a)}) for the symmetry of $\hat{H}$, however, is replaced by the requirement,\footnote{This condition appears because the weight $|x|$ in the scalar product (\ref{scalar_product}) splits the integral, $<\psi ,\hat{H} \phi> = \int_{a^-}^{a^+} \psi^{*} |x| \,(\hat{H}\phi) \,\mathrm{d} x =
-\int_{a^-}^{0} \psi^{*}  x (\hat{H}\phi) dx + \int_{0}^{a^+} \psi^{*} x (\hat{H}\phi) dx$.}
\begin{equation}
\psi^{*}(a^-)\phi(a^-) - \psi^{*}(0^{-})\phi (0^{-})-\psi^{*}(0^{+})\phi (0^{+}) +\psi^{*}(a^+)\phi(a^+) =0 . \label{new_condition}
\end{equation}

The Hamiltonian operator $\hat{H}= -i \hbar x^{-1}\partial_x$ is singular at $x=0$ and hence, the wave function is not defined there.  The correct definition of the domain is not the continuous interval  $(a^-,a^+)$, but rather  $(a^-,0)\cup (0,a^+)$, and the wavefunction $\psi(x)$ is allowed to be discontinuous at $x=0$. Therefore, we look for solutions that are everywhere bounded but not necessarily continuous at $x=0$, where they can present a (finite) discontinuity. This means, in particular, that (\ref{new_condition}) must be interpreted as two separate statements,
\begin{eqnarray}
\psi^{*}(a^-)\phi(a^-) =\psi^{*}(0^{-})\phi (0^{-}) \quad \mbox{and} \quad \quad \psi^{*}( a^+)\phi(a^+) =\psi^{*}(0^{+})\phi (0^{+}),
\end{eqnarray}
and the results of the previous section be can expected to hold for both intervals $(a^-,0)$ and $(0,a^+)$ separately.  

The self-adjoint condition for $\hat{H}$ must be respected in both domains, so the wavefunctions must satisfy the following boundary conditions 
\begin{equation}
\psi(a^-)=e^{i \theta^-}\psi(0^-), \quad \mbox{and} \quad \psi(a^+)=e^{i \theta^+}\psi(0^+)\, .
\end{equation} 
This in turn implies that the Hilbert space splits into two subspaces $\mathcal{H}_{\pm}$
\begin{eqnarray}
\mathcal{H}_{-}&=&\{ \phi(x) \in L^{2}((a^-,0), |x| dx) :  \phi(a^-)=e^{i \theta^-}\phi(0^{-})  \} \\
\label{L2-} 
\mathcal{H}_{+}&=&\{ \phi(x) \in L^{2}((0,a^+),|x| dx) :  \phi(a^+)=e^{i \theta^+}\phi(0^{+}) \}\, , 
\label{L2+}
\end{eqnarray}
where in general $\phi(0^-) \neq \phi(0^+)$.  This shows that the quantum problem in a region that extends across a degenerate surface reduces to the previous case on the disjoint sets $(a^-,0)$ and $(0,a^+)$, and the Hilbert
space splits into a direct sum
\begin{equation}
\mathcal{H} = \mathcal{H}_{-}  \oplus \mathcal{H}_{+}, 
\label{hilbertspaceasimm}
\end{equation}
where $\mathcal{H}_{-}=L^{2}((a^-,0), |x| dx)$ and $\mathcal{H}_{+}=L^{2}((0,a^+), |x| dx)$ are mutually orthogonal projections of $L^{2}((a^-,a^+), |x| dx)$ on the intervals $(a^-,0)$ and $(0,a^+)$. These projections can be implemented through the action of the operator $\hat{P}$, defined as $\hat{P}: f(x) \mapsto sgn[x]\cdot f(x)$, and
\begin{equation}
\mathcal{H}_{\pm}=\frac{1}{2}(\hat{P} \pm 1) \cdot L^{2}((a^-,a^+), |x| dx) \,.
\end{equation}
In this splitting, the support of each function space is restricted to either one side or the other, and the wave functions are
\begin{equation}
\psi= \left( \begin{array}{c}
\psi^+(x) \\
\psi^-(x) 
\end{array} \right)= \psi^+(x)  \oplus \psi^-(x) ,
\end{equation}
where
\begin{eqnarray}
\psi^-(x) &=&\left\{ \begin{array}{l} \frac{\sqrt{2}}{a^-} \exp \left[ \frac{i}{2\hbar \nu} E_n^- x^2 \right], \,\,\,
a^-<x<0 \\
0\, , \qquad \qquad \qquad \qquad \,\, 0<x<a^+\end{array} \right. \\
\psi^+(x)&=&\left\{ \begin{array}{c} 0\, , \qquad \qquad \qquad \qquad  \,\, a^-<x<0 \\
\frac{\sqrt{2}}{a^+} \exp \left[ \frac{i}{2\hbar \nu} E_n^+ x^2 \right], \,\,\,
0<x<a^+ \\
\end{array} \right. 
\end{eqnarray}
These are admissible solutions of the Schr\"odinger equation,
\begin{equation}
\hat{H} \psi^{\pm} = E^{\pm} \psi^{\pm} \label{Schr}
\end{equation} 
where the eigenvalues $E^{\pm}$ are found to be
\begin{equation}
E_n^{\pm}= (2n \pi + \theta^{\pm})\frac{2\hbar \nu}{(a^{\pm})^2} .
\label{Epm}
\end{equation}
The Hamiltonian splits into a block-diagonal form, each block having its own spectrum $\{E^{\pm}_n\}$. The complete energy spectrum is the union of the two spectra
\begin{equation}
\{ E_n\}=\{E^+_n\} \cup \{E^-_n\}\, .
\end{equation}
Some eigenvalues could have a matching one on the other side, \textit{i.e.} $E^+_{n} = E^-_m$, then
\begin{equation}
n=\left(\frac{a^+}{a^-}\right)^2 m + \kappa \, ,
\label{n=n(m)}
\end{equation}
where $\kappa = [(a^+)^2\theta^- -(a^-)^2\theta^+]/[2\pi (a^-)^2]$. As in the previous case, changing the energy spectrum by a constant corresponds to an equal shift in the phases of all wave functions, $\theta^{\pm}\rightarrow \theta^{\pm}+ \delta
\theta$ with no observable effects. This freedom can be used to set $\kappa=0$, so that the ground states on both sides ($n=0=m$) have the same energy. In that case, we can distinguish three possibilities:

 $\bullet$ If $(a^+/a^-)^2$ is a generic irrational number, the two spectra have only one common eigenvalue --a doubly degenerate ground state--

$\bullet$ If $(a^+/a^-)^2$ takes a rational value, there are some doubly degenerate eigenstates and the rest are nondegenerate

$\bullet$ In the extreme case a symmetric domain, $a^+=a^-$, all states are doubly degenerate.

\noindent
Then the general time-dependent solution reads
\begin{equation}
\Psi(x,t) = \left\{
\begin{array}{rl}
 &\sum c_m^{-} \frac{\sqrt{2}}{a^-} \exp \left[ 2\pi im ( x^2 - 2\nu t)(a^-)^{-2} \right] , \quad x \in (a^-,0)\\
\\ 
&\sum c_n^{+} \frac{\sqrt{2}}{a^+} \exp \left[2\pi in (x^2 - 2\nu t)(a^+)^{-2} \right], \quad x  \in (0,a^+)
\end{array} \right.\label{separetesols}
\end{equation}
where the coefficients are given as before
\begin{eqnarray*}
c_m^{-}     &=& < \Psi_{n}(x,t) | \Psi (x,t)> = \int_{a^-}^{0} \Psi_{n}^{*}(x,t) |x| \Psi (x,t) dx\\
c_n^{+}     &=& < \Psi_{m}(x,t) | \Psi (x,t)> = \int_{0}^{a^+}  \Psi_{m}^{*}(x,t) |x| \Psi (x,t) dx
\end{eqnarray*}

Note that there is no overlap between wavefunctions with support on opposite sides of the degeneracy surface. Consequently, a wave packet initially prepared in the region $x<0$ will never evolve into $x>0$, and vice-versa. This is in complete agreement with the classical behaviour of the system, in which the orbits on one side of the degeneracy surface never reach the other side. In
other words, there is neither classical nor quantum flow across the degeneracy surface.

The term $|x|$ in the probability density $\rho(x,t)=|\Psi(x,t)|^2 |x|$ reflects the role of the degeneracy as a singularity of the probability flow, where particle states are created or annihilated. In fact, the quantum mechanical probability density satisfies a continuity equation with a sink (or source) at the degeneracy,
\begin{equation}
\partial_{t}\rho+\partial_{x}J=\sigma\, , \label{deg_cont_eq}
\end{equation}
where
\begin{eqnarray}
J&=& \nu \, \mbox{sgn}(x) \, |\Psi(x,t)|^2 \\ 
\sigma &=& 2 \nu \, \delta(x) \, |\Psi(x,t)|^2. \label{sink}
\end{eqnarray}
Here the direction of the flow is determined by the sign of $\nu$ and the degeneracy at $x=0$ acts as sink ($\nu<0$) or source ($\nu>0$) of states. In any case, there is no net flux of the wave packet across the degeneracy.

\section{Discussion}  

\textbf{1.} Although the discussion here has been restricted to a rather simple case in which the configuration space corresponds to the entire real axis, it is easy to see that the conclusions do not change radically if the configuration space is compact ($x\in S^1$).  The only modifications introduced by the topology of the phase space are essentially two:\\ 
i) The restrictions imposed by the Poincar\'e-Hopf theorem that relates the degree of the singularities in the Hamiltonian flow and the Euler characteristic of the phase space manifold \cite{Milnor} \\
ii) The fact that the orbits reverse orientation at the degeneracy surfaces \cite{Thesis-deMicheli}. This second restriction implies that closed orbits on a compact manifold must intercept an even number of degeneracy surfaces.

\textbf{2.} The orthogonality of the Hilbert spaces on different sides of a degeneracy surface is not affected by the fact that some eigenvalues might accidentally match --\textit{e.g.}, if (\ref{n=n(m)}) holds. Even in the extreme case, in which all eigenvalues are identical (for $a^-=a^+$), the eigenstates supported on different sides are still orthogonal. In that case, the states could also be arranged into a basis of symmetric and antisymmetric wavefunctions, and not supported only on one side or the other. Such a basis of parity eigenstates may be convenient since the parity operator commutes with the Hamiltonian.

\textbf{3.}  The conclusion of no quantum tunnelling obtained in the above approach is consistent with other methods like Dirac's Hamiltonian approach, Feynman's path integral, or the old Bohr-Sommerfeld quantization condition.

\textbf{4.}  Lovelock Lagrangians that generalize the Einstein-Hilbert theory for dimensions higher than four, as well as Chern-Simons theories in dimensions five or more, are widely studied models for the potential description of spacetime and fundamental interactions at high energies. However, these actions are well known to present degeneracies \cite{TZ,BGH,HTrZ,MTrZ,Z}, and should be regarded, therefore, as describing a host of different physical systems. Depending on the initial conditions where the system starts, the evolution may take the system towards a degeneracy surface, where the number of degrees of freedom is reduced and the resulting field theory has an effective dynamics that corresponds to fewer dimensions, a sort of dynamical dimensional reduction as discussed in \cite{HTrZ}.

\subsection{Summary}    

Let us summarize our results:

$\bullet$ Two types of degenerate systems can be distinguished: those whose orbits never intersect the surfaces of degeneracy (reducible systems) and the rest (irreducible ones). In the first case, there exists an action principle which yields the same dynamics everywhere in phase space (with the possible exception of the degenerate surfaces, where the orbits may not exist). 
Irreducible systems on the other hand cannot be described by a nondegenerate action principle.

$\bullet$  The quantum mechanical description is obtained by the canonical substitution as in the standard Schr\"odinger picture, c.f., Eqs. (\ref{commutator}-\ref{operator_prescriptions}). The only difference is that the degeneracy of the symplectic form becomes the singular set of the quantum Hamiltonian operator. Since the singular points must be removed from the domain of the Hamiltonian operator, for consistency they should also be removed from the domain of the wave functions. This means that the Hilbert space must allow for wave functions that can be discontinuous on the degenerate surfaces.

$\bullet$  Allowing discontinuous wave functions implies that the solutions have support restricted to a single region bounded by a degenerate surface $\Sigma$. This has been realized by a Hilbert space that splits into a direct sum of orthogonal subspaces. Generalizing this conclusion, we see that since $\Sigma$ defines a collection of Jordan curves, the Hilbert space that describes a quantum degenerate system must be of the form 
\begin{equation}
\mathcal{H}=\oplus_{i=1}^N \mathcal{H}_i ,
\end{equation}
where each Hilbert subspace describes the dynamics of a subsystem supported in only one nondegenerate domain.

$\bullet$  The physical consequence of this is that there is no overlap between wavefunctions on different nondegenerate regions, and therefore no tunnelling across surfaces of degeneracy. This is in complete analogy with the classical picture of degenerate systems.

\subsection{Open questions}  

In \cite{STrZ} it was shown how a coupled system composed by subsystems, a degenerate and a nondegenerate one, evolve in time. The same question can be asked about the corresponding quantum system. The answer of this question will
be the subject of a forthcoming article \cite{Thesis-deMicheli,FJ}. 

A particularly interesting question is to understand how the present discussion extends to field theories. In particular, this would allow deciding whether the Chern-Simons systems in five spacetime dimensions or more are reducible or not. Although the notion of orbit in a field theory is only formally defined, some of the essential features of the distinction between reducible and irreducible systems can be applied. It might be conjectured that CS systems for$D\geq 5$ are generically irreducible (they cannot be replaced by a non-degenerate action principle), it is far from obvious how to settle this question.

Another, even more difficult question is how does a CS system behaves quantum mechanically.

\section*{\Large Acknowledgments}
We would like to thank Mokhtar Hassa\"ine, Olivera Mi\v{s}kovi\'c, Jorge Alfaro, Sergio Cacciatori, Livio Pizzocchero and Renato Nobili for many useful comments and enlightening discussions. This work was supported by Fondecyt grants \# 1110102, 1100328, 1100755, 1100328, and by Conicyt grant \textit{Southern Theoretical Physics Laboratory, ACT-91}. The Centro de Estudios Cient\'{\i}ficos (CECS) is funded by the Chilean Government through the Centers of Excellence Base Financing Program of Conicyt.

\subsection*{Appendix. Local coordinate transformations}   
\textbf{A. Darboux's Theorem}\\
 
We have seen that it is impossible to set $f(\mathbf{z})=1$ globally by a coordinate change but, is it possible to do it \textit{within an open nondegenerate} neighborhood? Is it possible to find appropriate coordinates, within each nondegenerate domain, so the dynamical equations look nondegenerate? 

Let us consider a degenerate system given as in (\ref{genericsystem}),
\begin{equation}
f(z)\dot{z}^i =\epsilon^{ij}\,\frac{\partial A_0(z)}{\partial z^j} , \qquad \epsilon^{12}=-\epsilon^{21}=1. \label{degsystem}
\end{equation}
In terms of new coordinates $\xi^i(z)$, this equation reads
\begin{eqnarray}
\dot{\xi}^a =\frac{1}{f}\frac{\partial\xi^a}{\partial z^i} \frac{\partial\xi^b}{\partial z^j} \epsilon^{ij}\frac{\partial
A_0(\xi)}{\partial\xi^b} = \frac{1}{f}\Big{\vert}\frac{\partial\xi}{\partial z}\Big{\vert} \epsilon^{ab} \frac{\partial A_0(\xi)}{\partial\xi^b} \,,
\label{xi-dot}
\end{eqnarray}
which reduces to (\ref{genericsystem}), provided 
\begin{equation}
\det (\partial \xi^i/\partial z^k)= f, \quad \quad \mbox{and} \quad\quad A_0(\xi)=A_0(z(\xi))\, .
\label{xi}
\end{equation}
There are certainly many choices of coordinates $\xi$ that satisfy conditions (\ref{xi}), so one concludes that there are many regular autonomous systems that have the same dynamics as a degenerate one within a nondegenerate region. Therefore, degenerate dynamical systems can always be reduced to non-degenerate ones in an open neighbourhood that does not include degeneracy surfaces. This explains why textbooks on differential equations never discuss degenerate systems.\\ 

\textbf{B. Time reparametrizations}   \\

The fact that the shape of the orbit is independent of $f$ suggests that a change of time parameter may yield an evolution that could be reproduced by a nondegenerate dynamical system. If that is the case, then a new time parameter $\tau(t)$ should exist such that
\begin{equation}
\frac{dt}{d\tau}=  \frac{1}{f({z}^{1},{z}^{2})} \label{t-tau} \,.
\end{equation}
This relation could be integrated if the trajectory $z^i(t)$ is known,
\begin{equation}
\tau(t;z_0) = \int_{t_0}^t{f(z^1(t'),z^2(t')) dt'} \, .
\end{equation}
This relation, however is not a redefinition of the time parameter for the entire dynamical system, but for each individual orbit. Moreover, the reparametrization $\tau=\tau(t)$ fails precisely at the degeneracy points, where $f$ changes sign. This highlights the fact that the degenerate surfaces that intersect the classical trajectories are starting or ending points of orbits.


\end{document}